# Molecular Beam Epitaxy of thin HfTe$_2$ semimetal films

S. Aminalragia-Giamini[*,1,3], J. Marquez-Velasco[1,4], P. Tsipas[1], D. Tsoutsou[1], G. Renaud[2], A. Dimoulas[1,5]

*1. National Center for Scientific Research "DEMOKRITOS", Patriarchou Grigoriou & Neapoleos 27, 15310, Agia Paraskevi, Athens, Greece*

*2. Univ. Grenoble Alpes and CEA, INAC-MEM, F-38000 Grenoble, France*

*3. University of Athens, Physics Department, Section of Solid State Physics, Panepistimiopolis, GR 15684 Zografos, Athens, Greece*

*4. Department of Physics, National Technical University of Athens, Heroon Polytechniou 9, GR 15780 Zografos, Athens, Greece*

*5. Chair of Excellence, Univ. Grenoble Alpes and CEA, INAC-MEM, F-38000 Grenoble, France*

Abstract

Epitaxial thin films of 1T-HfTe$_2$ semimetal are grown by MBE on AlN(0001) substrates. The measured *in-plane* lattice parameter indicates an unstrained film which is also azimuthally aligned with the AlN substrate, albeit with an *in-plane* mosaic spread, as it would be expected for van der Waals epitaxy. Angle resolved photoemission spectroscopy combined with first principles electronic band structure calculations show steep linearly dispersing conduction and valence bands which cross near the Brillouin zone center, providing evidence that HfTe$_2$ /AlN is an epitaxial topological Dirac semimetal.

[*]E-mail: s.agiamini@inn.demokritos.gr   tel.number: +302106503325

**Introduction**

Atomically thin materials of the metal-dichalcogenide family ($MX_2$, M=Metal, X=S, Se, Te) crystallized in a stable 2D form have been the subject of intense investigation due to their versatile physical properties [1]. The molybdenum and tungsten sulfides and selenides are among the best studied 2D $MX_2$ materials. Stabilized in the hexagonal 2H prismatic crystal structure they are semiconductors with sizable gap between 1 and 2 eV making them suitable for nano and optoelectronic applications. The corresponding transition metal ditellurides ($MoTe_2$ and $WTe_2$) have a more complex structural and electronic behavior. In $WTe_2$, due to strong W-W intermetallic bonds, there is a lattice distortion which stabilizes this material in the orthorhombic ($T_d$) semimetallic phase [2]. $MoTe_2$ in its most stable 2H phase is a semiconductor and it has already been grown in its epitaxial form by MBE [3]. However, it can also be prepared in the distorted monoclinic (1T') [4] or low temperature (<260K) orthorhombic ($T_d$) [5] metastable phases which are both semimetallic. Following predictions $WTe_2$, [6] and $T_d$-$MoTe_2$ [7] have been experimentally proven [8-10] very recently to be type II Weyl semimetals (WSM). WSM are new forms of topological quantum matter and have brought much excitement due to analogies with high energy physics and the prospect for radically new applications in electronics and spintronics [11,12]. A related class of topologically non-trivial matter are the 3D Dirac semimetals (DSM) which are two overlapping WSM in momentum and energy space. The DSMs, known also as "3D graphene" are formed by conduction and valence band crossing at some point of the Brillouin zone with a linear dispersion in all three directions in k-space. Most of experimentally proven Weyl semimetals (TaAs, NbAs) [13,14] and Dirac semimetals ($Na_3Bi$, $Cd_3As_2$) [15,16] are bulk materials and are synthesized by bulk CVT methodologies. On the other hand metal (di)tellurides and other 2D and layered semimetals can be synthesized in stable atomically thin film form either by exfoliation or by epitaxial growth on suitable large area substrates. This permits the investigation of possible interesting nontrivial topology at the very low thickness regime, eventually down to a single layer, and permits the fabrication of devices in compatibility with the scaling trends in current nanoelectronics. $T_d$-MoTe$_2$ WSM and ZrTe$_5$ Dirac semimetal [17] are two examples of 2D layered materials with non-trivial topology but so far only bulk crystals have been synthesized by CVT. It is not clear whether thin epitaxial films can be obtained from these materials, especially for the metastable orthorhombic WSM phase of MoTe$_2$.

HfTe$_2$ and ZrTe$_2$ formed in the stable 1T phase can be alternative 2D di-tellurides whose properties remain unexplored. Early work [18-20] provided evidence through electronic transport measurements that HfTe$_2$ is a semimetal but detailed information about its electronic band structure, especially for thin films, is lacking. Since the bulk Hf-Te compounds have been synthesized by CVT [18-22] and theoretically predicted [23, 24] to appear in several stoichiometries (HfTe$_2$, HfTe$_3$, HfTe$_5$), control in the desired stoichiometry by film growth methodologies could be challenging and needs to be investigated.

Here we present the first epitaxial HfTe$_2$ thin films, grown by MBE on AlN(0001)/Si(111) substrates. The epitaxial quality, the surface morphology, the microstructure and the electronic band structure are investigated in detail. Comparison with first principles calculations suggests the presence of non-trivial topology.

**Methods**

The HfTe$_2$ layers with several thicknesses from 2 to 10 monolayers (ML) are grown on AlN(0001)/Si(111) substrates using Molecular Beam Epitaxy (MBE) in an ultrahigh vacuum system with a base pressure of 10$^{-10}$ Torr. A two-step process is employed where in the first step high-purity Hf and Te are evaporated using e-gun and effusion cell respectively, at a substrate temperature of 530 °C. In the second step, the sample is annealed *in-situ* at 650 °C for 15 minutes to improve the crystalline quality. The ratio of the evaporation rates $R_{Te}/R_{Hf}$ is kept high (~15/1) in order to ensure sufficient incorporation of Te in the compound. Following growth the samples were characterized *in-situ* by Reflection High Energy Electron Diffraction (RHEED), Angle-Resolved Photoemission Spectroscopy (ARPES) and Scanning Tunneling Microscopy (STM). ARPES measurements were made employing a 2D CCD detector with a He I excitation source at 21.22 eV. STM measurements were carried out with an Omicron Large Sample-STM at room temperature. Subsequently, HfTe$_2$ samples were capped with a thin aluminium film (5nm) to prevent oxidation and investigated by Grazing Incidence X-ray Diffraction (GIXRD) in a UHV-MBE-CVD diffractometer [25] at the European Synchrotron Radiation Facility (ESRF) using two different X-ray energies/wavelengths of 11 keV (1.127 Å) and 9.5 keV (1.305 Å). The incident beam, of energy resolution 10$^{-4}$, had a very small (5 x10$^{-5}$ rad) divergence parallel to the surface. The incident angle was set at 0.22°, which is slightly below the critical angle for total external reflection perpendicular to the surface, in order to enhance the thin film signal while minimizing the background from the AlN buffer layer.

The characterization results discussed below reveal that under these growth conditions the compound is produced by MBE in the 1T-$HfTe_2$ phase. No evidence of $HfTe_3$ or $HfTe_5$ phases is obtained. Finally, Density Functional Theory (DFT) calculations were performed using the Vienna Ab initio Simulation Package [26] and projector-augmented waves. The generalized-gradient approximation (GGA) [27] with Perdew–Burke–Ernzerhof (PBE) [28] parameterization was used as exchange correlation functional. The semi-empirical DFT-D3 Grimme's method [29] was applied to include van der Waals corrections. The HfTe2 layers were optimized using a threshold of $1 \times 10^{-5}$ eV/Å for the Hellmann-Feynman force as a criterion. The plane wave kinetic energy cutoff was set at 500 eV, using the Monkhorst-Pack scheme [27] employing a 21x21x1 k-point grid. Calculation of the band structure with and without spin-orbit coupling (SOC) were performed using a k-mesh of 31 k-points per symmetric line along MΓKM and ΓA directions.

**Results and Discussion**

*Electron Diffraction*

Reflection high energy electron diffraction (RHEED) patterns for the AlN substrate and HfTe$_2$ epilayer are given in Fig. 1. The sharp streaky pattern of the epilayer indicates a well ordered and flat HfTe$_2$ surface. Using the AlN RHEED pattern as reference and knowing that the lattice constant of wurtzite AlN is 3.11 Å, the lattice constant of the HfTe$_2$ is estimated to be 3.94Å. This is very close to the reported [18-20] value of bulk, free standing HfTe$_2$, which is equal to 3.95Å. Despite the large lattice mismatch (27%) it is remarkable that there is a good azimuthal alignment of the two materials indicating epitaxial growth without strain in agreement with van der Waals heteroepitaxy. The faint additional streaks along the [1-100] azimuth indicates that a small amount of 30 º or 90 º rotated domains may exist in the epilayer. A more detailed picture of the alignment and the domain microstructure can be obtained from high resolution X-Ray data given below.

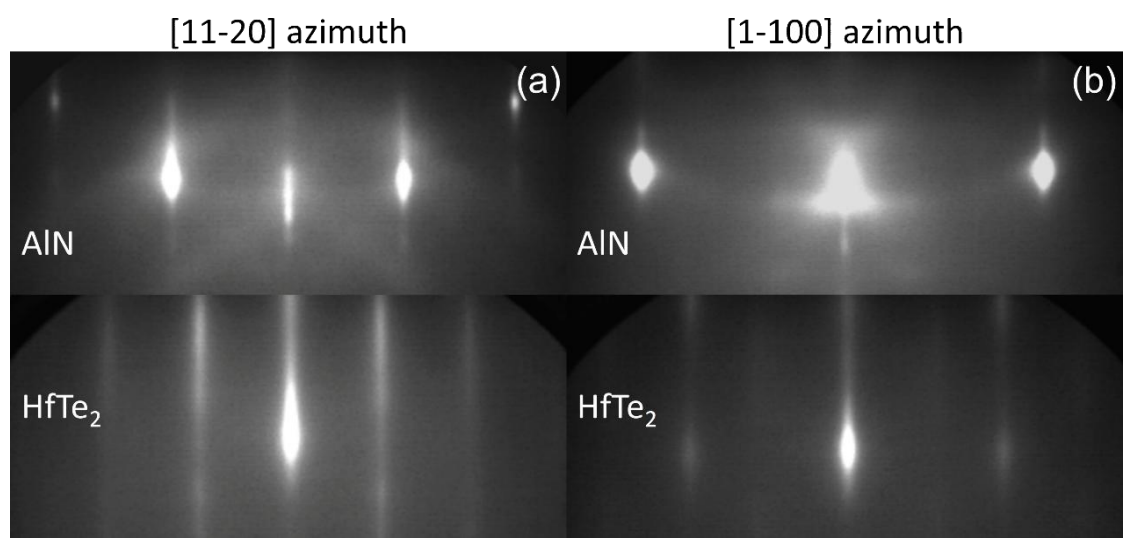

Figure 1. RHEED pattern of HfTe$_2$ grown on AlN(0001)/Si(111) substrates along (**a**) [11–20] azimuth and (**b**) [1-100] azimuth. The two materials show very good alignment

despite the lattice mismatch. The sharp streaky patterns indicate good surface ordering and flat surfaces of HfTe$_2$.

*Surface structure and morphology*

The surface of a HfTe$_2$ epitaxial film is investigated by *in-situ* room temperature UHV-STM. Figure 2 shows a large area scan of 500x500 nm$^2$ (Fig. 2a) and a linescan across a layer step (Fig. 2b). The surface morphology (Fig. 2a) shows that the film is grown in a two-dimensional island form in such a way that a layer is almost fully formed before the next one starts to grow. Small darker areas are depressions where the layer is not fully grown. From the linescan, it can be inferred that the height difference between two layers is 6.79Å. This value is very close to the expected [18-20] vertical lattice parameter equal to 6.67Å along the [0001] axis of HfTe$_2$,

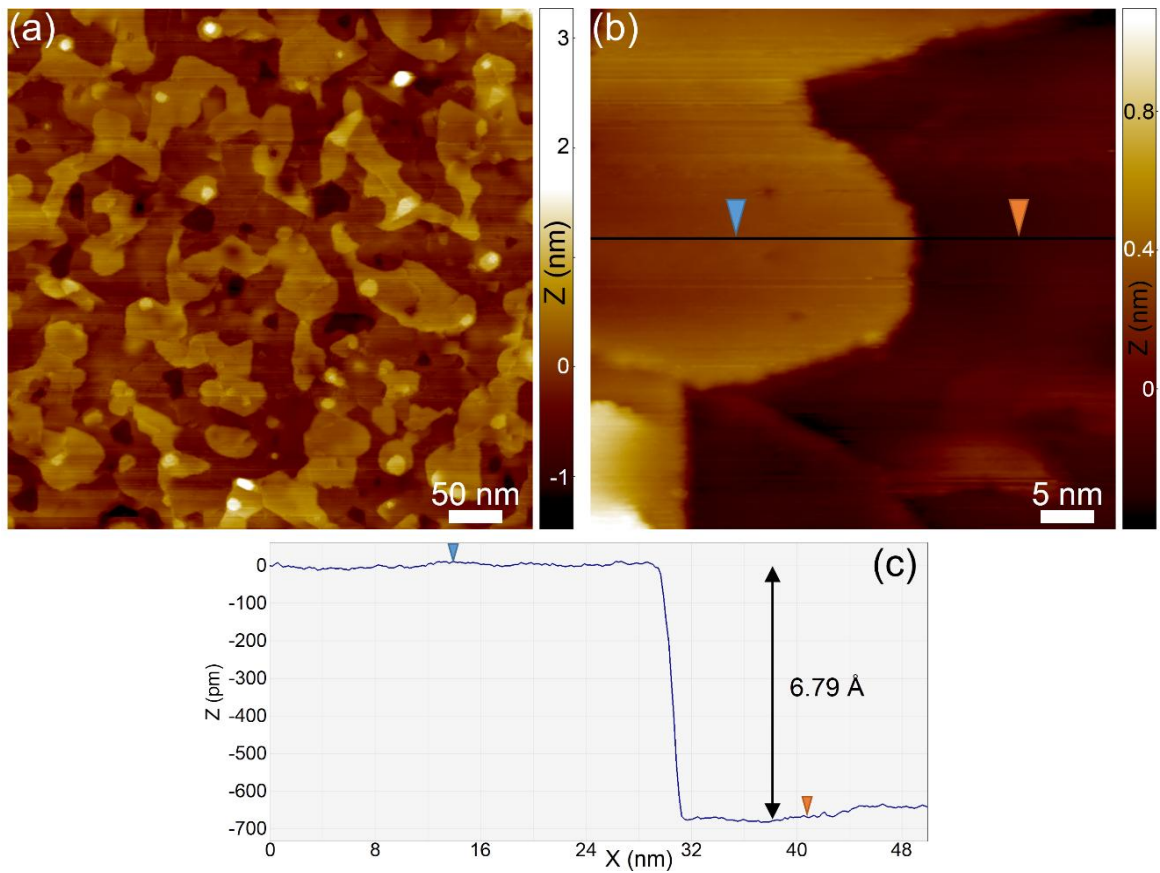

Figure 2. Large area scans of HfTe$_2$ grown on AlN/Si(111). (**a**) A scan area of 500x500 nm$^2$ shows a two-dimensional island growth mode where each layer grows after the previous. (**b**) The step between two layers and (**c**) the linescan from (b) measuring a vertical distance of 6.79Å, very close to the theoretical of 6.67Å.

A high resolution image and a line scan are given in Fig. 3 along with the theoretically derived structure of 1T-HfTe$_2$ using DFT. The image shows the top Te atoms in a hexagonal arrangement with an in-plane lattice parameter of 3.97Å in good agreement with the experimental values [18-20] as well as the theoretically estimated value of 3.94Å derived by our DFT calculations shown in Fig. 3c and 3d.

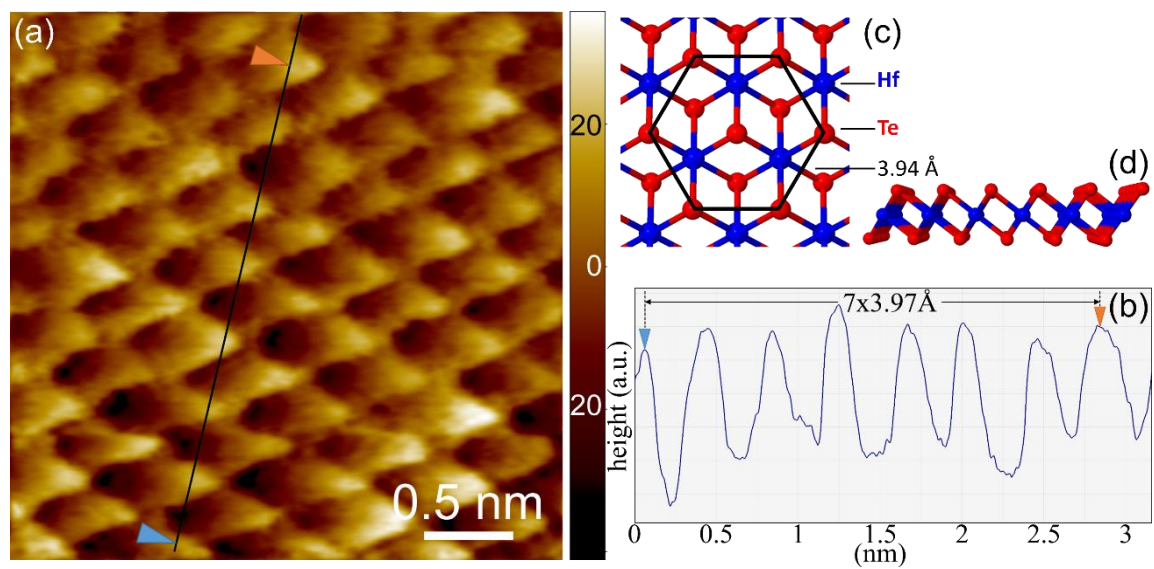

Figure 3. (**a**) High resolution scan of HfTe$_2$ grown on AlN/Si(111) in a 3x3 nm$^2$ area. (**b**) The linescan from (a) showing a measured in-plane lattice constant of 3.97Å very close to the reported value of 3.95Å, (**c**) and (**d**) atom arrangement of the 1T-HfTe$_2$ structure calculated with DFT showing the structure from a top and side view respectively.

*Epitaxial quality and microstructure*

Figure 4a shows a reciprocal space map (RSM) covering one sixth of the in-plane reciprocal space produced by GIXRD measurements. The sample was capped with Al to prevent oxidation and the thickness was estimated by GIXRD to be 2-3 ML. As shown in the Supplementary Information, precise radial scans were performed along the high symmetry directions as well as measurements along the rods of scattering perpendicular to the surface on all measurable Bragg peaks/rings. Rocking scans (see Fig. 4b) were also performed on all of them. On the in-plane RSM the diffraction from three hexagonal lattices is clearly apparent, the sharp but hardly visible peaks of the Si(111) substrate, the peaks from the 200 nm-thick hexagonal AlN layer and those of the $HfTe_2$ thin layer. No other feature is visible, except a wide ring attributed to the (2-20) Bragg reflection of the aluminium capping layer. The positions of these peaks (together with out-off plane ones, see Supplementary Information), yield the epitaxial relationships: Si[1-10](111)//AlN[10-10](0001)// $HfTe_2$[10-10](0001)//Al[1-10](111). These results are in fairly good agreement with the RHEED images and they further prove that the material is azimuthally aligned with the AlN substrate despite the large lattice mismatch, consistent with van der Waals epitaxial growth. It should be noted that no peaks are detected at 30° positions within the sensitivity limits of the technique indicating that the quantity of 30° rotated domains that may exist (see also discussion of RHEED data, Fig. 1) is negligibly small.

The in-plane lattice parameter of the $HfTe_2$ layer is deduced by fitting the positions of the corresponding Bragg peaks along radial scans, which yields a lattice constant of 3.967Å. This again is in very good agreement with the reported value for $HfTe_2$ [18-20] as well as with what was estimated in this work by the analysis of RHEED and STM images and the theoretical value derived from our DFT calculations.

Furthermore, the Full Width at Half Maximum (FWHM) of the HfTe$_2$ peaks along radial scans are also used to estimate the average *in-plane* domain size *D* defining areas of pristine single crystalline quality which is found to be between 10 and 20 nm. Moreover, the in-plane mosaic spreads were deduced from rocking scan measurements across the AlN and HfTe$_2$ Bragg peaks (see Fig. 4b) yielding a mosaic spread of 11 ± 0.5° for HfTe$_2$, significantly larger than the mosaicity of AlN which is only 0.7º. Finally, the out-off plane thickness and structure, and in particular the stacking sequence, were determined by simulating the rods of scattering of the HfTe$_2$ layer (Supplementary). The position of the out-off plane allowed and forbidden Bragg peaks demonstrates that the HfTe$_2$ layer is of 1T structure, as opposed to 2H [30-32] (Supplementary). The structural parameters deduced from the different X-ray measurements are summarized in Table 1.

| Sample | AlN / HfTe$_2$ 2-3ML / Al capping |
|---|---|
| In-plane AlN lattice parameter | 3.121 ± 0.001 Å |
| In-plane AlN mosaic spread | 0.69° ± 0.03° |
|  |  |
| In-plane HfTe$_2$ lattice parameter | 3.967 ± 0.002 Å |
| In-plane HfTe$_2$ mosaic spread | 11 ± 0.5° |
| In-plane HfTe$_2$ domain size | 10 < D < 20 nm |
| Out-off plane HfTe$_2$ lattice parameter | 6.66 ± 0.01 Å |
|  |  |
| In-plane Al lattice parameter | 4.052 ±0.002 Å |
| In-plane Al mosaic spread | ~12.7° |

Table 1. Summary of results for the calculated lattice parameter and mosaic spread of a AlN/HfTe$_2$/Al(capping) sample for the three materials. The average domain size is also shown for HfTe$_2$, calculated to be between 10 and 20 nm.

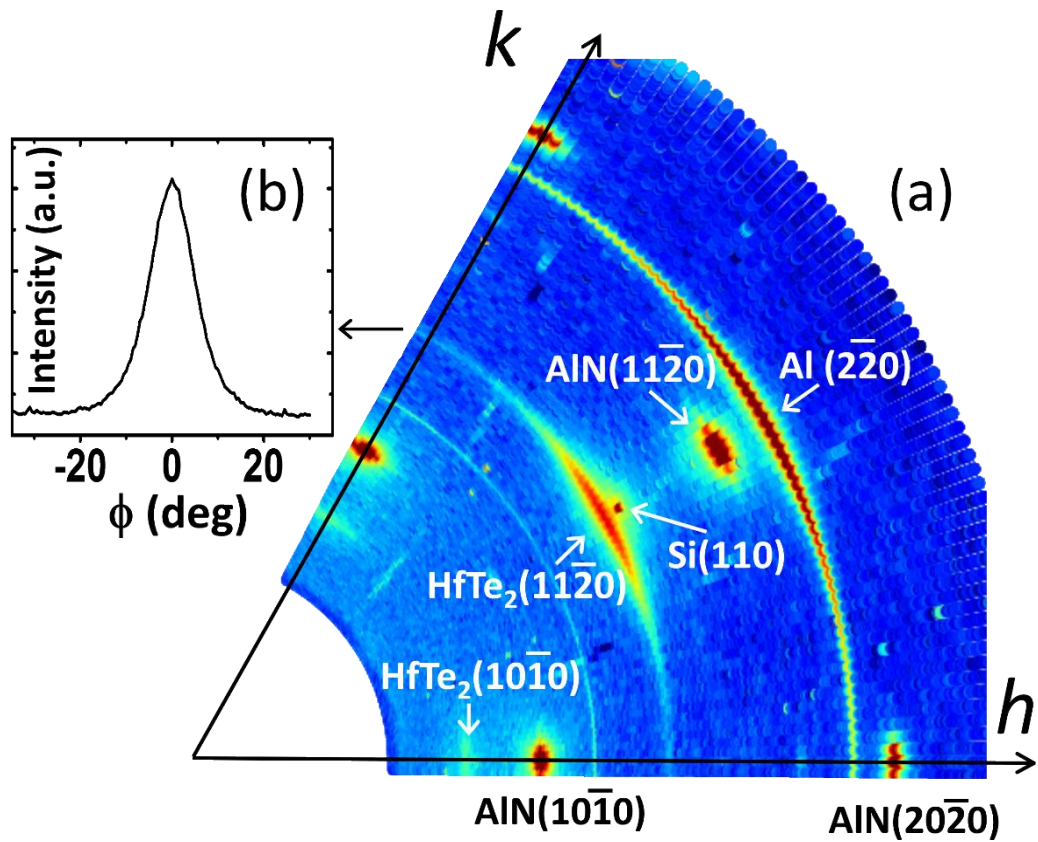

Figure 4. (a) In-plane reciprocal space map (RSM) of a 2-3ML thick HfTe$_2$ sample, measured by rocking the sample over 70° at increasing values of the in-plane Si(111) surface reciprocal lattice units $h$ and $k$ with increments of 0.01. The color scale is logarithmic, the highest (red) intensity being $10^6$ ph/seconds and the background ~100 ph/s. The HfTe$_2$, AlN and Al Bragg peaks are labelled using bulk Miller indexed. The Si(110) peak is labelled using surface reciprocal lattice units. (b) Azimuthal rocking scan across the HfTe$_2$(11-20) reflection indicating a mosaic spread of 11° FWHM.

*Electronic band structure*

The electronic band structure of 1T-HfTe$_2$ has been investigated by first principles calculations (DFT) using PBE potentials with van der Waals corrections, taking into account SOC. The results for the bulk along the MΓKM direction in the Brillouin zone are shown in Fig. 5a, while results for 1, 4 and 10 ML-thick 1T-HfTe$_2$ free standing films as well as 1ML 1T-HfTe$_2$ on AlN are presented in the supplementary for comparison. One of the main band structure characteristics is that the p-orbital Te valence band and d-orbital Hf conduction band nearly cross at the center of the Brillouin zone (Γ) with only a mini-gap opening, of ~14meV (see supplementary). The near crossing point is found to be 0.4 eV above the Fermi level suggesting a significant energy overlapping between the valence band at Γ and the conduction band at M, which is a characteristic of a semimetallic material. A more detailed examination of the band dispersion in Fig. 5(b) reveals that the near crossing occurs close to the Γ-point at $k_z$ = 0.045Å$^{-1}$ along the ΓA direction of the Brillouin zone, showing a striking similarity with the (near) crossing of the bands in Na$_3$Bi topological 3D Dirac semimetal (supplementary) which produces a pair of Dirac cones experimentally observed [15]. This similarity suggests that HfTe$_2$ could be a topological Dirac semimetal candidate that is grown epitaxially in stable 2D form, thus presenting an advantage over the well-known Na$_3$Bi or Cd$_3$As$_2$ Dirac semimetals appearing only in bulk form [15,16]. Furthermore, by examining in detail the orbital character of the conduction and valence band in Fig. 5b, a possible band inversion is indicated implying non-trivial topology.

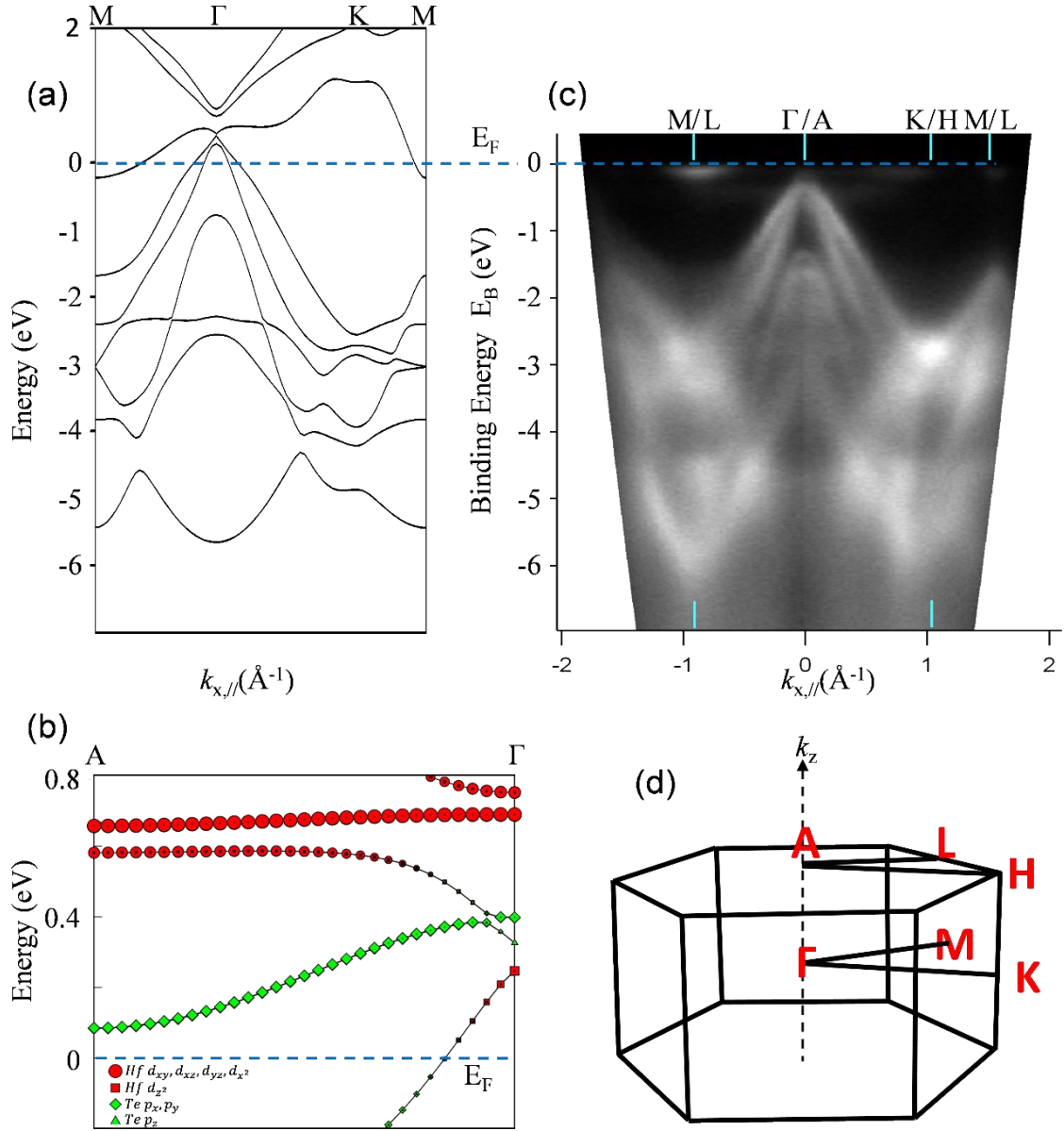

Figure 5. (**a**) DFT band structure calculation of bulk 1T-HfTe$_2$ with spin-orbit coupling along the high symmetry MΓKM direction. (**b**) Detailed view of the DFT calculation along the A-Γ direction where the orbital character of the bands shows inversion and the near-crossing close to the Γ point with a mini-gap of 14 meV. (**c**) ARPES valence band imaging using He I excitation of 21.22 eV along the high symmetry directions K/H-Γ/A-M/L of the Brillouin zone of a HfTe$_2$ sample (**d**) The hexagonal Brillouin zone of 1T-HfTe$_2$ showing high-symmetry points.

The HfTe$_2$ electronic band structure of a 4ML film is imaged along high symmetry crystallographic directions in the surface Brillouin zone by using *in-situ* angle resolved photoelectron spectroscopy (ARPES) and the results are presented in Fig. 5c. In general there is a fairly good agreement between theory (Fig. 5a, and supplementary) and experiment (Fig. 5c), especially regarding the steep variation of the valence band showing a maximum at Γ and the appearance of conduction band minima at the M points as predicted by theory. The experimental results indicate a semimetal but, compared to theory (Fig. 5a), there is a smaller energy overlap between conduction and valence bands (Fig. 5c). This is mainly due to an apparent shift of the valence band at Γ to lower energies in such a way that its maximum is located exactly on the Fermi level, instead of being 0.4 eV above it as predicted by theory (Fig. 5a and Supplementary). This could be interpreted as a result of charge exchange with the substrate or unintentional defect-related doping which could raise the Fermi level closer to the apex of the cone where VB and CB nearly touch each other at the Γ point. However in this case we would expect that the Fermi level is positioned well above the conduction band minima at M, which is not supported by the ARPES images (Fig. 5c) since the Fermi level is clearly located near the bottom of the CB at M. Although misfit or thermal strain could modify the band structure potentially explaining the discrepancy with theory near Γ, the presence of strain is not supported by the structural characterization results all of which suggest a relaxed (unstrained) film with lattice parameter near the bulk value despite evidence of small lattice inhomogeinities from GIXRD.

A faint band is observed (Fig. 5c) connecting the valence band near Γ with the conduction band at M. Since it does not appear in our DFT calculations of the bulk bands (Fig. 5a), we attribute this feature to a surface band with possible non-trivial topology in connection to the band inversion predicted in Fig. 5b. It should be noted that similar (topological) surface states connecting electron and hole pockets have been observed recently [33-35] in the orthorhombic ($T_d$) phase of the WSM $MoTe_2$.

Finally, the faint feature appearing at $k_{x,//} = 0.8 Å^{-1}$ along ΓK in Fig. 5c, not present in theoretical calculations (Fig. 5a), originates from the CB minima at the M points in neighboring BZs as better explained in the discussion of the Fermi surface below.

The measured Fermi surface in a portion of the k-space is shown in the $k_x$-$k_y$ constant energy contour plots (CECP) in Fig. 6a, 6b. With the help of the Brillouin zone diagram in Fig. 6c, we identify the electron pockets around the M points which are clearly observed in both CECP (Fig. 6a and 6b); a narrower hole pocket at the Γ point is also visible. The two electron pockets around M, seen in the CECP in Fig. 6b, have a finite spread and produce the faint signal in the E-$k_{x,//}$ image of Fig. 7a which is not expected from theoretical simulations (Fig. 5a).

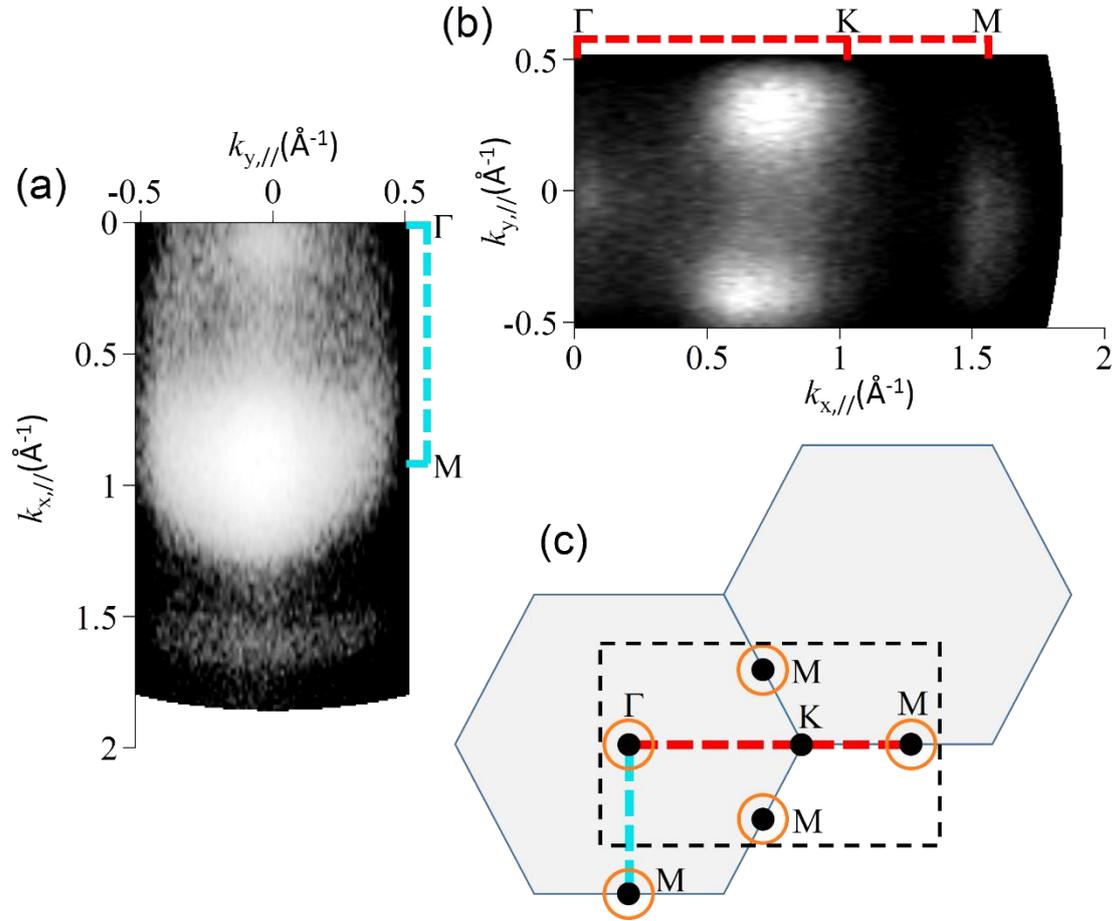

Figure 6. (**a**) constant energy $k_x$-$k_y$ contour plots of the Fermi surface along the high symmetry ΓM direction where the electron pocket and the top of the hole pocket are visible at the M and Γ points respectively. (**b**) constant energy $k_x$-$k_y$ contour plots of the Fermi surface along the high symmetry ΓK direction where the electron pockets at the adjacent and neighboring M points are visible. (**c**) Schematic of Brillouin zones and the Γ, M and K points with the imaging window of ARPES.

We further investigate the band dispersion at the BZ center close to the Fermi level as shown in Fig. 7. The E-$k_{//}$ dispersion (Fig. 7a) and the CECPs (Fig. 7b) show a structure consisting of two cone-like bands with hexagonal distortions, one within another. The outer band exhibits a parabolic behavior with a maximum ~ 0.2 eV below $E_F$. On the other hand the inner band shows a steep linear dispersion crossing the outer band at $E_b$

~-0.4 eV and forming a cone with its apex located exactly at the Fermi level. The linear variation of the inner cone indicates Dirac-like dispersion which is characteristic of massless particles. The Fermi velocity $v_F$ is calculated to be ~0.76 $10^6$ m/sec which is by 25% smaller compared to that of graphene. The behavior is similar to the dispersion reported in 3D topological Dirac semimetal $Na_3Bi$. Taking into account the similarities with $Na_3Bi$ [15] and the predicted near crossing behavior in Figs. 5a and 5b as discussed above, it is anticipated here that the linear dispersion near Γ in Fig. 7 provides evidence that $HfTe_2$/AlN is an epitaxial topological Dirac semimetal. Unlike the case of $Na_3Bi$ and $Cr_3As_2$ Dirac semimetals which exist in bulk form, $HfTe_2$ is a 2D material and can be scaled down to a single monolayer either by epitaxial growth or by exfoliation thus providing an excellent opportunity to investigate the transition of topological 3D Dirac semimetal properties to the 2D limit.

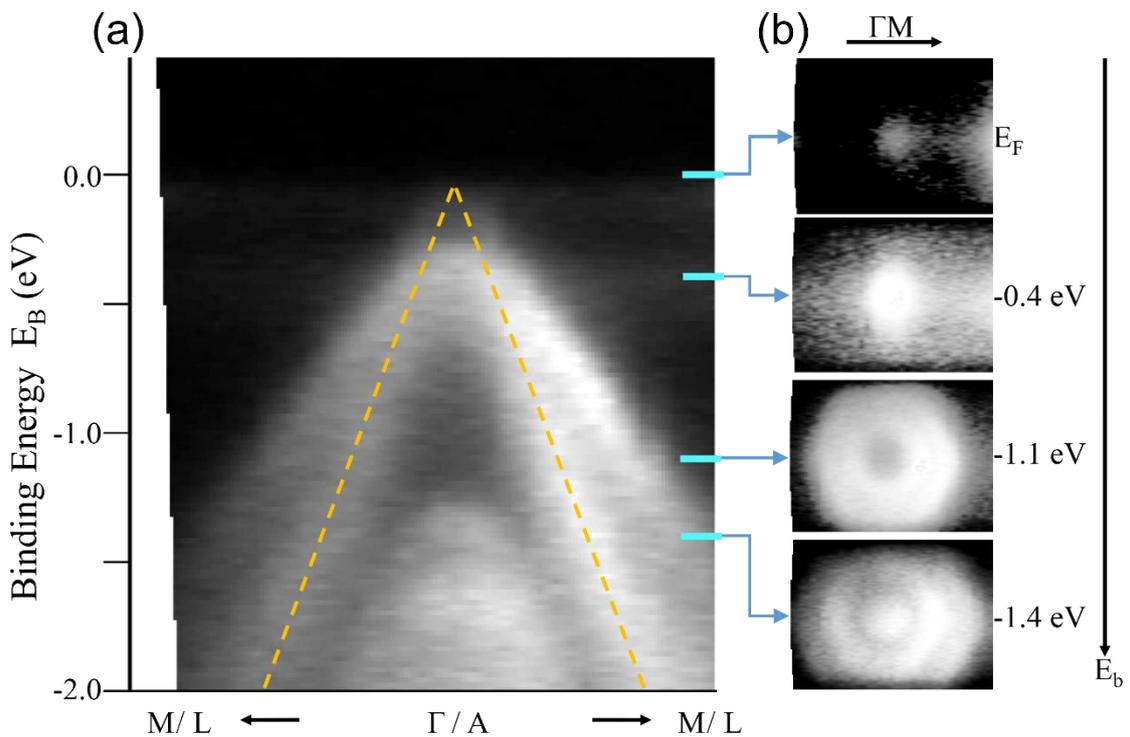

Figure 7. (**a**) Detail of the band structure around the Γ point. The inner bands form a Dirac-like cone which touches the Fermi level at its apex (**b**) constant energy $k_x$-$k_y$

contour plots at different energies. The apex of the inner cone is evident at the Fermi level and at lower energies both cones show hexagonal distortions.

**Conclusions**

In this work we present the first epitaxial HfTe$_2$ semimetal grown on AlN(0001) substrates by MBE. We show that the material adopts a 2D island growth mode with a very good surface ordering. HfTe$_2$ is azimuthally aligned with the substrate, albeit with a mosaicity of 11º FWHM. The measured lattice constant agrees very well with that of bulk 1T-HfTe$_2$ indicating that HfTe$_2$ grows on the substrate with no average strain resembling a free standing film. The electronic band structure is imaged revealing a Dirac cone-like behavior of the valence band near the center of the Brillouin zone with the cone apex lying on the Fermi level. A direct comparison with first principle calculations, indicates that HfTe$_2$/AlN is a possible epitaxially grown topological Dirac semimetal.

Acknowledgements: We acknowledge financial support from ERC Advanced Grant SMARTGATE-**291260**